\documentclass[reprint,showpacs,amsmath,amssymb,aps,pre]{revtex4-1}
\usepackage{dcolumn}
\usepackage{bm}
\usepackage[dvipdfmx]{graphicx}
\usepackage[dvipdfmx]{hyperref}
\usepackage[mathlines]{lineno}

\begin{document}

\title{Energy polarization and energy pumping in Rice-Mele chains}
\author{Kiminori Hattori, Kodai Ishikawa, and Yuma Kaneko}
\affiliation{Department of Systems Innovation, Graduate School of Engineering Science, Osaka University, Toyonaka, Osaka 560-8531, Japan}
\date{\today}

\begin{abstract}	
The one-dimensional Rice-Mele lattice consisting of two-site unit cells enables topological charge pumping at low enough temperatures.
We show here that energy pumping is feasible in this system.
We analyze energy polarization using the second-quantization formalism and connect it to intercell energy current by the relevant continuity equation.
Following this formulation, we numerically evaluate the energy current pumped in an adiabatic cycle when the system parameters vary periodically and slowly with time.
Unlike the pumped charge, the pumped energy is entirely temperature-independent at half filling.
The global Berry phase for all bands and the related topological phase transitions during the pumping cycle account for the energy pumping.
The present study paves the way for manipulating energy current without electric and thermal biases.
\end{abstract}

\maketitle

\section{Introduction}
\label{sec:1}

Topological quantum states of matter have attracted immense attention in recent years since they produce fundamentally new physical phenomena and have potential applications in novel devices.
Historically, the concept of topological matter originates from the theoretical formulation of the integer quantum Hall effect by Thouless et al. \cite{ref:1}.
They showed that the first Chern number, a topological invariant quantized at integers, accounts for the robust quantization of Hall conductivity in two dimensions (2D).
The subsequent discovery of time-reversal-symmetric topological insulators revealed the ubiquitous existence of topological materials, leading to the widespread study of topological aspects in insulators, semimetals, and superconductors \cite{ref:2,ref:3,ref:4}.
Recently, the objects of study have expanded beyond electronic systems to include non-electronic counterparts such as mechanic \cite{ref:5,ref:6}, acoustic \cite{ref:7,ref:8}, photonic \cite{ref:9,ref:10,ref:11}, and ultracold atomic systems \cite{ref:12,ref:13,ref:14,ref:15,ref:16}.

The well-known Su-Schrieffer-Heeger (SSH) model forms a prototype for topological insulators, which consists of a 1D bipartite lattice with two sublattice sites in each unit cell \cite{ref:17}.
Since this model preserves chiral symmetry, its band topology is described by a topological invariant termed the winding number \cite{ref:4}.
The SSH model is generalized to the Rice-Mele (RM) model by including a staggered sublattice potential \cite{ref:18}.
It is well known that adiabatic charge pumping is enabled in the RM model when the system parameters are modulated periodically and slowly with time \cite{ref:19,ref:20,ref:21}.
The charge pump transports topologically quantized charge between neighboring cells without a driving electric field.
The resulting adiabatic current is analogous to the dissipationless quantum Hall current.
As demonstrated by Thouless, the pumped charge per cycle is explicitly described by the Berry curvature in time-momentum space and the associated Chern number \cite{ref:22,ref:23,ref:24}.
The charge pumping also provides a firm foundation for the modern theory of electric polarization, which is connected to the Berry phase across the Brillouin zone \cite{ref:19,ref:20,ref:21}.
The topological charge pump has been realized in the recent experiments utilizing ultracold atoms \cite{ref:15,ref:16}.

The present study shows that energy pumping is also realizable in the RM model.
The energy pump under consideration operates at a single driving frequency similarly to a conventional charge pump and differs from the multi-frequency or non-monochromatic energy pump proposed in the literature \cite{ref:25,ref:26,ref:27,ref:28,ref:29}.
The paper is organized as follows.
In Sec. \ref{sec:2}, we introduce a winding number that classifies topological phases in the RM model.
In this model, chiral symmetry is broken, and each band is topologically trivial.
Nonetheless, the global Berry phase for all bands \cite{ref:30} leads to the $\mathbb{Z}$ invariant.
We also show that this classification is valid for generic 1D systems regardless of symmetry.
In Sec. \ref{sec:3}, we describe charge and energy polarizations in the second-quantization formalism and relate them to the relevant intercell currents in terms of current continuity equations.
On the basis of these results, in Sec. \ref{sec:4}, we formulate instantaneous currents flowing in an arbitrary adiabatic cycle at finite temperatures.
Following these formulations, in Sec. \ref{sec:5}, numerical calculations are used to compare charge and energy currents produced in a specific pumping cycle at half filling.
The pumped charge is quantized at low enough temperatures while vanishing in the high-temperature limit.
Unlike the pumped charge, the pumped energy is entirely temperature-independent, and quantitatively explained in terms of the global Berry phase for all bands and the related topological phase transitions during the pumping cycle.
Finally, Sec. \ref{sec:6} provides a summary.

\section{Model and Berry Phase}
\label{sec:2}

Throughout this paper, we shall work in units where $e = \hbar = k_{\text{B}} = 1$.
We consider the RM model for noninteracting spinless fermions.
The relevant single-particle Hamiltonian is written as $\mathcal{H} = {\sum _{jj'\alpha \beta }}\left| {j,\alpha } \right\rangle H_{jj'}^{\alpha \beta }\left\langle {j',\beta } \right|$ in real space, where $j \in \{ 1,2, \cdots ,N\} $ denotes the lattice position of the two-site unit cell, $\alpha ,\beta \in \{ A,B\} $ represents the sublattice degree of freedom in each cell, and $\left| {j,\alpha } \right\rangle $ is the basis ket at each site.
The Hamiltonian matrix $H_{jj'}^{\alpha \beta }$ is explicitly expressed as
\begin{eqnarray}
\label{eq:1} 
H_{jj'}^{AB} & = & H_{j'j}^{BA} = v{\delta _{jj'}} + w{\delta _{j,j' + 1}},\\
\label{eq:2}
H_{jj'}^{AA} & = & - H_{jj'}^{BB} = m{\delta _{jj'}},
\end{eqnarray}
where $v$ ($w$) denotes the intracell (intercell) hopping energy, and $m$ describes the staggered sublattice potential.
For simplicity, we assume that $v$ and $w$ are nonnegative in the following.
Note that if $m = 0$, the RM model is reduced to the SSH model.
In momentum space, the Hamiltonian is formulated as $H(k) = {N^{ - 1}}{\sum _{jj'}}{H_{jj'}}{e^{ - ik(j - j')}} = {\mathbf{d}}(k) \cdot \boldsymbol{\sigma} $, where $\boldsymbol{\sigma} = ({\sigma _x},{\sigma _y},{\sigma _z})$ is the Pauli vector, and the 3D vector ${\mathbf{d}} = ({d_x},{d_y},{d_z})$ is composed of ${d_x} = v + w \cos k$, ${d_y} = w \sin k$, and ${d_z} = m$.
The eigenequation $H(k)\left| {{u_n}(k)} \right\rangle = {\varepsilon _n}(k)\left| {{u_n}(k)} \right\rangle $ is solved to be ${\varepsilon _n} = nd$ and
\begin{equation*}
\left| {{u_n}} \right\rangle = \frac{1}{{\sqrt {2d(d - n{d_z})} }}\left( {\begin{array}{*{20}{c}}
 {n({d_x} - i{d_y})} \\ 
 {d - n{d_z}} 
\end{array}} \right) ,
\end{equation*}
where $n = \pm 1$ denotes the band index, and $d = \sqrt {d_x^2 + d_y^2 + d_z^2} $.
The corresponding Berry connection is given by
\begin{equation}
\label{eq:3}
A_k^{nn} = i\left\langle {{u_n}} \right|\frac{\partial }{{\partial k}}\left| {{u_n}} \right\rangle = \frac{{{d_x}\frac{{\partial {d_y}}}{{\partial k}} - {d_y}\frac{{\partial {d_x}}}{{\partial k}}}}{{2d(d - n{d_z})}},
\end{equation}
for band $n$.
In terms of $A_k^{nn}$, the Berry phase across the 1D Brillouin zone, i.e., the Zak phase, is expressed as ${\nu _n} = {(2\pi )^{ - 1}}\smallint _0^{2\pi }dkA_k^{nn}$ in units of $2\pi $.
This quantity is not quantized unless $m = 0$.
Instead, summing over two bands, we obtain
\begin{equation}
\label{eq:4}
\sum\limits_n {A_k^{nn}} = \frac{{{d_x}\frac{{\partial {d_y}}}{{\partial k}} - {d_y}\frac{{\partial {d_x}}}{{\partial k}}}}{{d_x^2 + d_y^2}} = \frac{{\partial \phi }}{{\partial k}} ,
\end{equation}
where $\phi = \tan ^{-1} ({d_y}/{d_x})$.
Hence, the total Berry phase
\begin{equation}
\label{eq:5}
\nu = \sum\limits_n {{\nu _n}} = \frac{1}{{2\pi }}\int_0^{2\pi } {dk\frac{{\partial \phi }}{{\partial k}}} ,
\end{equation}
amounts to a winding number that describes how often the 2D vector $({d_x},{d_y})$ winds about the origin as $k$ varies over the entire Brillouin zone \cite{ref:31}, and is simply accounted for by the projection onto $xy$ plane of $\mathbf{d}$ vector.
(Although a non-Hermitian RM model is discussed in this literature, the total Berry phase derived there is equivalent to that in the Hermitian case.)
The winding number $\nu $ is a topological invariant and is generically integer quantized.
For the present model, one finds $\nu = 0$ for $v > w$ and $\nu = 1$ for $v < w$.
Notably, $\nu $ is independent of $m$ and insusceptible to a small variation in the parameter space $(v,w)$.
The SSH model preserves chiral symmetry.
In this case, ${\nu _n} = \nu /2$ is quantized equally for two bands at a half-integer.

\begin{figure}
\centering
\includegraphics[width=\linewidth]{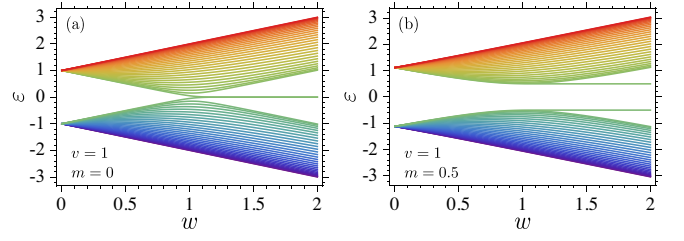}
\caption{(Color online) Eigenenergies $\varepsilon $ calculated as a function of $w$ for the open RM chain consisting of $N = 30$ unit cells. Two panels compare the numerical results for (a) $m = 0$ and (b) $m = 0.5$. In these figures, $v$ is taken as the energy unit ($v = 1$).}
\label{fig:1}
\end{figure}

In a finite RM chain with open ends, two ingap edge modes emerge in the nontrivial phase $v < w$, as shown in Fig. \ref{fig:1}.
Thus, the presence or absence of edge modes correlates to the bulk topology defined by the winding number $\nu $.
In this sense, the bulk-boundary correspondence holds for the RM model.
It is also worth noting that the band gap remains open in the RM model with $m \ne 0$, and the topological phase transition occurs without closing the gap \cite{ref:31}.
This is distinct from the feature seen in the SSH model with $m = 0$, where the gap closes at the transition point $v = w$.
However, the phase transition irrelevant to gap closing does not contradict topological band theory, since $\nu $ is not a single-band invariant but the multiband one that characterizes all energy states.
For a long enough RM chain, the eigenfunctions are analytically derived for the two edge modes to be $\left| L \right\rangle \propto {\sum _j}\left| {j,A} \right\rangle {( - v/w)^{j - 1}}$ and $\left| R \right\rangle \propto {\sum _j}\left| {j,B} \right\rangle {( - v/w)^{N - j}}$.
These solutions fulfill the eigenequations $\mathcal{H}\left| L \right\rangle = m\left| L \right\rangle $ and $\mathcal{H}\left| R \right\rangle = - m\left| R \right\rangle $.
The existence condition of edge modes is $v < w$.
This criterion is identical to that for $\nu = 1$.

The above argument for the specific 1D model is generalizable by considering a non-Abelian Berry connection.
The Berry connection $A_k^{nn'} = i\left\langle {{u_n}} \right|{\partial _k}\left| {{u_{n'}}} \right\rangle $ between two bands $n$ and $n'$ is expressed compactly as ${A_k} = i{U^{ - 1}}{\partial _k}U$ in matrix notation, where $U$ is the unitary matrix composed of the full set of eigenvectors.
Since $\det U$ is a unit complex number expressed as ${e^{ - i\phi }}$, we derive $\operatorname{Tr} {A_k} = {\partial _k}\phi $ from the identity $\operatorname{Tr} {U^{ - 1}}{\partial _k}U = {\partial _k}\ln \det U$.
Therefore,
\begin{equation}
\label{eq:6}
\nu = \frac{1}{{2\pi }}\int_0^{2\pi } {dk \operatorname{Tr} {A_k}} = \frac{1}{{2\pi }}\int_0^{2\pi } {dk\frac{{\partial \phi }}{{\partial k}}} ,
\end{equation}
reads as a winding number.
The Berry connection matrix ${A_k}$ involves all bands in the system.
For this reason, we refer to $\nu $ as the global Berry phase for all bands \cite{ref:30} to distinguish it from the single-band Berry phase ${\nu _n}$.
Equation (\ref{eq:6}) is valid regardless of symmetry.
Equation (\ref{eq:5}) exemplifies this for the RM model where chiral symmetry is broken.
As shown by these equations, the global Berry phase (in units of $2\pi $) is equivalent to the winding number, and hence cannot be defined modulo 1.
Instead, its gauge-invariant form is given by ${\sum _n}({\nu _n} \mod 1)$ for nondegenerate bands. However, such a technical formulation is not essential in the following discussion.

\section{Polarization and Current}
\label{sec:3}

In the second-quantization formalism, the real-space Hamiltonian is written as $\mathcal{H} = {\sum _{jj'\alpha \beta }}c_{j\alpha }^\dag H_{jj'}^{\alpha \beta }{c_{j'\beta }}$, where $c_{j\alpha }^\dag $ (${c_{j\alpha }}$) is the fermionic creation (annihilation) operator for a particle at sublattice site $\alpha $ in cell $j$.
This Hamiltonian is diagonalized as $\mathcal{H} = {\sum _{nk}}{\varepsilon _n}(k)c_{nk}^\dag {c_{nk}}$, where ${c_{nk}} = {\sum _{j\alpha }}{c_{j\alpha }}b_{nk\alpha }^*(j)$ is the fermion operator in momentum space and ${b_{nk\alpha }}(j) = {N^{ - 1/2}}\left\langle {\alpha }
 \mathrel{\left | {\vphantom {\alpha {{u_n}(k)}}}
 \right. \kern-\nulldelimiterspace}
 {{{u_n}(k)}} \right\rangle {e^{ikj}}$ is the Bloch function.
These formulations are helpful in dealing with polarizations and currents in the RM model, as shown below.

In this formalism, charge density (the number of particles in a cell $j$) is expressed as
\begin{equation}
\label{eq:7}
{\rho _j} = \sum\limits_\alpha {c_{j\alpha }^\dag {c_{j\alpha }}} .
\end{equation}
In terms of the equation of motion ${\dot c_{j\alpha }} = - i{\sum _{j'\beta }}H_{jj'}^{\alpha \beta }{c_{j'\beta }}$, the operator ${\rho _j}$ follows the continuity equation ${\dot \rho _j} + {\sum _{j'}}J_{jj'}^{(\rho )} = 0$, where
\begin{equation}
\label{eq:8}
J_{jj'}^{(\rho )} = i\sum\limits_{\alpha \beta } {H_{jj'}^{\alpha \beta }(c_{j\alpha }^\dag {c_{j'\beta }} - c_{j'\beta }^\dag {c_{j\alpha }})} ,
\end{equation}
describes the charge current flowing from cell $j$ to cell $j'$.
It is easily shown from the symmetry $H_{jj'}^{\alpha \beta } = H_{j'j}^{\beta \alpha }$ that $J_{jj'}^{(\rho )} = - J_{j'j}^{(\rho )}$ and hence $J_{jj}^{(\rho )} = 0$.
By definition, it is also evident that $J_{jj'}^{(\rho )} = 0$ for $\left| {j - j'} \right| > 1$.
These constraints reduce the continuity equation to ${\dot \rho _j} + J_{j,j + 1}^{(\rho )} + J_{j,j - 1}^{(\rho )} = 0$.
In terms of ${\rho _j}$, charge polarization is formulated as
\begin{equation}
\label{eq:9}
{P_\rho } = \frac{1}{N}\sum\limits_j {{\rho _j}j} .
\end{equation}
Following the above argument, one finds ${\dot P_\rho } = {N^{ - 1}}{\sum _j}J_{j,j + 1}^{(\rho )}$.
Considering that $J_{j,j + 1}^{(\rho )}$ is independent of $j$ because of translational symmetry, we finally reach a simple relation
\begin{equation}
\label{eq:10}
{\dot P_\rho } = J_{j,j + 1}^{(\rho )} ,
\end{equation}
at the operator level, showing that ${\dot P_\rho }$ accounts for intercell charge transport.

The analogous formulation is derived for energy polarization and energy current.
Using the equation of motion, the real-space Hamiltonian is rewritten in the form $\mathcal{H} = {\sum _j}{h_j}$.
Here,
\begin{equation}
\label{eq:11}
{h_j} = \frac{i}{2}\sum\limits_\alpha {c_{j\alpha }^\dag {{\dot c}_{j\alpha }}} + h.c. ,
\end{equation}
represents energy density (the energy of particles in a cell $j$).
It is easy to show that the operator ${h_j}$ obeys the continuity equation ${\dot h_j} + {\sum _{j'}}J_{jj'}^{(h)} = 0$, where
\begin{equation}
\label{eq:12}
J_{jj'}^{(h)} = \frac{1}{2}\sum\limits_{\alpha \beta } {H_{jj'}^{\alpha \beta }(c_{j'\beta }^\dag {{\dot c}_{j\alpha }} - c_{j\alpha }^\dag {{\dot c}_{j'\beta }})} + h.c. ,
\end{equation}
describes the energy current flowing from cell $j$ to cell $j'$.
Energy polarization is definable as
\begin{equation}
\label{eq:13}
{P_h} = \frac{1}{N}\sum\limits_j {{h_j}j} .
\end{equation}
Equation (\ref{eq:13}) is structurally equivalent to Eq. (\ref{eq:9}) and hence naturally leads to
\begin{equation}
\label{eq:14}
{\dot P_h} = J_{j,j + 1}^{(h)} ,
\end{equation}
which relates energy polarization to intercell energy current.

Both charge and energy polarizations reflect Berry phases, as shown from their expectation values in thermal equilibrium, expressed as
\begin{eqnarray}
\label{eq:15}
\left\langle {{P_\rho }} \right\rangle & = & \frac{1}{{2\pi }}\sum\limits_n {\int_0^{2\pi } {dk{f_n}A_k^{nn}} } ,\\
\label{eq:16}
\left\langle {{P_h}} \right\rangle & = & \frac{1}{{2\pi }}\sum\limits_n {\int_0^{2\pi } {dk{\varepsilon _n}{f_n}A_k^{nn}} } ,
\end{eqnarray}
where ${f_n} = f({\varepsilon _n})$ and $f(\varepsilon ) = {[{e^{(\varepsilon - \mu )/\theta }} + 1]^{ - 1}}$ is the Fermi function.
In deriving the above formulas, we used the position matrix in the Bloch basis given by
\begin{equation*}
\sum\limits_{j\alpha } {b_{nk\alpha }^*(j){b_{n'k'\alpha }}(j)j} = [i\frac{\partial }{{\partial k}}{\delta _{nn'}} + A_k^{nn'}(k)]{\delta _{kk'}} .
\end{equation*}
Assuming a chemical potential $\mu$ lying in the gap, one finds
\begin{equation}
\label{eq:17}
\lim \limits_{\theta \to 0} \left\langle {{P_\rho }} \right\rangle = {\nu _ - } ,
\end{equation}
in the low-temperature limit.
The ground-state expectation value derived above is consistent with the modern theory of charge polarization in terms of Wannier states \cite{ref:19,ref:20,ref:21} and the Resta formulation using a momentum-translation operator \cite{ref:32,ref:33}.
In the high-temperature limit, we obtain
\begin{eqnarray}
\label{eq:18}
\lim \limits_{\theta \to \infty } \left\langle {{P_\rho }} \right\rangle & = & \frac{\nu }{2} ,\\
\label{eq:19}
\lim \limits_{\theta \to \infty } \left\langle {{P_h}} \right\rangle & = & \frac{{m\nu }}{2} ,
\end{eqnarray}
irrespective of $\mu$.
The latter is derived by noting that ${\sum _n}nA_k^{nn} = ({d_z}/d){\sum _n}A_k^{nn}$.

Before ending this section, we briefly summarize the expectation values of charge and energy densities.
They are generally expressed as $\left\langle {{\rho _j}} \right\rangle = {(2\pi )^{ - 1}}{\sum _n}\smallint _0^{2\pi }dk{f_n}$ and $\left\langle {{h_j}} \right\rangle = {(2\pi )^{ - 1}}{\sum _n}\smallint _0^{2\pi }dk{\varepsilon _n}{f_n}$, respectively.
Assuming $\mu = 0$, one finds ${f_ + } + {f_ - } = 1$ and ${f_ - } - {f_ + } = \tanh (d/2\theta )$.
Then, charge density satisfies $\left\langle {{\rho _j}} \right\rangle = 1$ irrespective of temperature, indicating that there is always one particle in each cell.
Conversely, energy density is reduced to $\left\langle {{h_j}} \right\rangle = - {(2\pi )^{ - 1}}\smallint _0^{2\pi }dk\gamma $, where $\gamma = d \tanh (d/2\theta )$.
Notably, $\left\langle {{h_j}} \right\rangle $ is always negative and vanishes in the high-temperature limit.

In the following part of this paper, we set $\mu = 0$.
In this case, the system remains half filled, and the energy current $\left\langle {J_{j,j + 1}^{(h)}} \right\rangle $ is equivalent to the heat current $\left\langle {J_{j,j + 1}^{(h)}} \right\rangle - \mu \left\langle {J_{j,j + 1}^{(\rho )}} \right\rangle $.

\section{Adiabatic Pumping}
\label{sec:4}

On the basis of the results derived in Sec. \ref{sec:3}, we next proceed to adiabatic pumping.
For brevity, we omit the notation $\left\langle \cdots \right\rangle $ in the following relations.
We may write charge and energy polarizations as
\begin{equation}
\label{eq:20}
P = \frac{1}{{2\pi }}\sum\limits_n {\int_0^{2\pi } {dk{g_n}A_k^{nn}} } ,
\end{equation}
in the unified notation, where ${g_n} = {f_n}$ for charge polarization, and ${g_n} = {\varepsilon _n}{f_n}$ for energy polarization.
Considering an adiabatic cycle when the system parameters such as $\{ v,w,m\} $ periodically and slowly vary with time $t$, then, Eq. (\ref{eq:20}) is formally valid, and the instantaneous charge and energy currents defined by $J = \dot P$ are formulated as
\begin{equation}
\label{eq:21}
J = \frac{1}{{2\pi }}\sum\limits_n {\int_0^{2\pi } {dk(\frac{{\partial {g_n}}}{{\partial t}}A_k^{nn} - \frac{{\partial {g_n}}}{{\partial k}}A_t^{nn} + {g_n}{B^{nn}})} } .
\end{equation}
Here, $A_x^{nn} = i\left\langle {{u_n}} \right|{\partial _x}\left| {{u_n}} \right\rangle $ is the Berry connection associated with a variation in $x \in \{ t,k\} $, and ${B^{nn}} = {\partial _t}A_k^{nn} - {\partial _k}A_t^{nn}$ represents the Berry curvature in $(t,k)$ space.
Thus, charge and energy pumped in a single period from $t = 0$ to $2\pi $ are given by
\begin{equation}
\label{eq:22}
Q = \int_0^{2\pi } {dtJ} ,
\end{equation}
which corresponds to the polarization difference $P(2\pi ) - P(0)$ between the initial and final states in a pumping cycle.
It is easy to see that $Q$ is gauge-invariant although $P$ and $J$ are gauge-dependent.

In the low-temperature limit (where $f_ + = 0$ and $f_ - = 1$), the pumped charge is reduced to
\begin{equation}
\label{eq:23}
\lim \limits_{\theta \to 0} {Q_\rho } = {C_ - } ,
\end{equation}
where ${C_n} = {(2\pi )^{ - 1}}\smallint _0^{2\pi }dt\smallint _0^{2\pi }dk{B^{nn}}$ denotes the Chern number.
For generic two-band models, ${C_n}$ is rewritten as ${C_n} = - n\lambda $, where
\begin{equation}
\label{eq:24}
\lambda = \frac{1}{{4\pi }}\int_0^{2\pi } {dt\int_0^{2\pi } {dk{\mathbf{\hat d}} \cdot (\frac{{\partial {\mathbf{\hat d}}}}{{\partial t}} \times \frac{{\partial {\mathbf{\hat d}}}}{{\partial k}})} } ,
\end{equation}
is a winding number that describes the number of times the 3D vector ${\mathbf{\hat d}} = {\mathbf{d}}/d$ wraps around the unit sphere when $(t,k)$ covers the 2D Brillouin zone.
In 1D, ${C_n}$ is also reducible to the winding of the single-band Berry phase, i.e., ${C_n} = \smallint _0^{2\pi }dt{\dot \nu _n}$.
In the high-temperature limit (where $f_ + = f_ - = 1/2$), one finds
\begin{equation}
\label{eq:25}
\lim \limits_{\theta \to \infty } {Q_\rho } = 0 ,
\end{equation}
since $\operatorname{Tr} B = 0$.

One can deal with energy pumping in the following ways.
Substituting ${g_n} = {\varepsilon _n}{f_n} = (nd - \gamma )/2$ into Eq. (\ref{eq:21}), energy current is decomposed into ${J_h} = {\bar J_h} + \Delta {J_h}$, where
\begin{eqnarray}
\label{eq:26}
{{\bar J}_h} & = & \frac{1}{{4\pi }}\int_0^{2\pi } {dk(\frac{{\partial {d_z}}}{{\partial t}}\frac{{\partial \phi }}{{\partial k}} - \frac{{\partial {d_z}}}{{\partial k}}\frac{{\partial \phi }}{{\partial t}})} ,\\
\label{eq:27}
\Delta {J_h} & = & - \frac{1}{{4\pi }}\int_0^{2\pi } {dk(\frac{{\partial \gamma }}{{\partial t}}\frac{{\partial \phi }}{{\partial k}} - \frac{{\partial \gamma }}{{\partial k}}\frac{{\partial \phi }}{{\partial t}})} ,
\end{eqnarray}
and ${\partial _x}\phi = \operatorname{Tr} {A_x}$.
The former is derived by noting that ${\sum _n}n{B^{nn}} = {\eta _t}{\partial _k}\phi - {\eta _k}{\partial _t}\phi $, where ${\eta _x} = (d{\partial _x}{d_z} - {d_z}{\partial _x}d)/{d^2}$.
In the RM model, ${d_z}$ ($ = m$) does not depend on $k$ so that ${\bar J_h} = \dot m\nu /2$.
Hence, we obtain
\begin{equation}
\label{eq:28}
{\bar Q_h} = \frac{1}{2}\int_0^{2\pi } {dt\frac{{dm}}{{dt}}\nu } .
\end{equation}
Note that ${\bar J_h}$ and ${\bar Q_h}$ are temperature-independent.
In the high-temperature limit, $\Delta {J_h}$ vanishes because of $\lim \limits_{\theta \to \infty } \gamma = 0$ so that
\begin{equation}
\label{eq:29}
\lim \limits_{\theta \to \infty } {Q_h} = {\bar Q_h} .
\end{equation}

The formulas given above are valid for an arbitrary pumping procedure.
Notably, charge pumping and energy pumping are different in the formulation.
The former is attributed to the nontrivial Chern number of the lower occupied band, whereas the latter is accounted for in terms of the global Berry phase for all bands and its time evolution.
Note that, unless $\nu $ varies temporally, ${\bar Q_h}$ vanishes because of periodicity of $m(t)$.

\section{Numerical Calculation}
\label{sec:5}

In the numerical calculations for charge pumping and energy pumping, we considered adiabatic variations of parameters such that $v(t) = 1 + \delta \cos t$, $w(t) = 1$, and $m(t) = \delta \sin t$.
In this pumping setup, ${C_ - } = \lambda = 1$ is verified.
At $t = 0$, the system is initialized in thermal equilibrium at a finite temperature $\theta $.
Assuming slow enough thermalization, the occupation number of each instantaneous eigenstate is invariant during the pumping cycle and is represented as ${f_n}(t,k) = {[{e^{{\varepsilon _n}(0,k)/\theta }} + 1]^{ - 1}}$, which satisfies ${\dot f_n} = 0$.
This assumption is used to numerically evaluate charge pumping and its temperature dependence for ultracold atoms \cite{ref:14}.
In contrast, fast enough thermalization allows an instantaneous equilibration for which ${f_n}(t,k) = {[{e^{{\varepsilon _n}(t,k)/\theta }} + 1]^{ - 1}}$.
Interestingly, the same result is numerically derived in these two opposing limits irrespective of temperature.
The instantaneous current $J(t)$ is computed using Eq. (\ref{eq:21}).
Its time-integration is denoted by $q(t) = \smallint _0^td\tau J(\tau )$ so that $Q = q(2\pi )$.
To further support our arguments, we also evaluated the equilibrium polarization $P$ by Eq. (\ref{eq:20}).

\begin{figure}
\centering
\includegraphics[width=\linewidth]{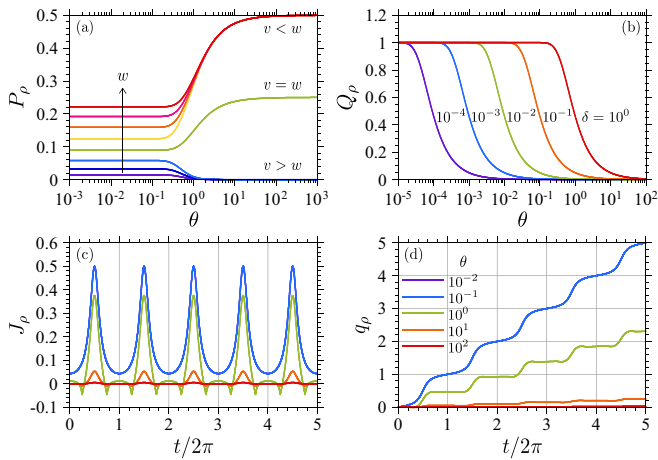}
\caption{(Color online) Numerical results for charge polarization and charge pumping. (a) Charge polarization ${P_\rho }$ as a function of $\theta $ for $w = 0.4,0.6, \cdots ,1.8$. In this calculation, we assumed $v = m = 1$. (b) Pumped charge ${Q_\rho }$ per cycle as a function of $\theta $ for $\delta = {10^{-4}},{10^{-3}}, \cdots ,{10^0}$. The lower two panels show (c) instantaneous charge current ${J_\rho }(t)$ and (d) its time-integration ${q_\rho }(t)$ in five pumping cycles for $\theta = {10^{-2}},{10^{-1}}, \cdots ,{10^2}$. In these calculations, we assumed $\delta = 1$.}
\label{fig:2}
\end{figure}

Figure \ref{fig:2} summarizes the numerical results for charge polarization ${P_\rho }$ and pumped charge ${Q_\rho }$.
In Fig. \ref{fig:2} (a), ${P_\rho }$ is shown as a function of $\theta $ for various $w$.
As $\theta \to \infty $, ${P_\rho }$ is saturated and becomes $1/2$ for $v < w$ and $0$ for $v > w$.
These results agree with those predicted by Eq. (\ref{eq:18}).
In Fig. \ref{fig:2} (b), ${Q_\rho }$ is shown as a function of $\theta $ for various $\delta $.
As expected from Eqs. (\ref{eq:23}) and (\ref{eq:25}), ${Q_\rho }$ is quantized at unity for $\theta << \delta $, whereas ${Q_\rho }$ varies as $\delta /2\theta $ for $\theta >> \delta $ and vanishes in the $\theta \to \infty $ limit.
These two regimes are separated by a critical temperature ${\theta _c} \approx \delta $ at which ${Q_\rho } = 1/2$.
This is a reasonable result since the RM chain opens a gap of size $2\delta $, which remains unchanged during the pumping cycle.
For $\theta << \delta $, only the lower band is occupied and contributes to charge pumping.
In Figs. \ref{fig:2} (c) and (d), ${J_\rho }(t)$ and ${q_\rho }(t)$ derived for various $\theta $ are summarized, respectively.
In the $\theta \to 0$ limit, ${q_\rho }(t)$ increases periodically by the amount of ${Q_\rho } = 1$ per cycle, while ${J_\rho }(t)$ and ${q_\rho }(t)$ vanish in the $\theta \to \infty $ limit.
These results do not contradict ${Q_\rho }$ shown in Fig. \ref{fig:2} (b).

\begin{figure}
\centering
\includegraphics[width=\linewidth]{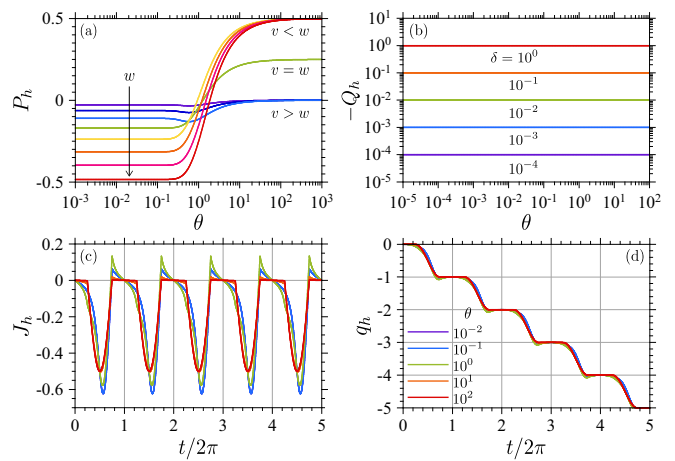}
\caption{(Color online) Numerical results for energy polarization and energy pumping. (a) Energy polarization ${P_h}$ as a function of $\theta $ for $w = 0.4,0.6, \cdots ,1.8$. In this calculation, we assumed $v = m = 1$. (b) Pumped energy ${Q_h}$ per cycle as a function of $\theta $ for $\delta  = {10^{-4}},{10^{-3}}, \cdots ,{10^0}$. The lower two panels show (c) instantaneous energy current ${J_h}(t)$ and (d) its time-integration ${q_h}(t)$ in five pumping cycles for $\theta = {10^{-2}},{10^{-1}}, \cdots ,{10^2}$. In these calculations, we assumed $\delta = 1$.}
\label{fig:3}
\end{figure}

Figure \ref{fig:3} depicts the numerical results for energy polarization ${P_h}$ and pumped energy ${Q_h}$.
In Fig. \ref{fig:3} (a), ${P_h}$ is shown as a function of $\theta $ for various $w$.
As seen in the figure, ${P_h}$ approaches $m/2$ for $v < w$ and $0$ for $v > w$ as $\theta \to \infty $.
These results are explained reasonably by Eq. (\ref{eq:19}).
In Fig. \ref{fig:3} (b), ${Q_h}$ is shown as a function of $\theta $ for various $\delta $.
As demonstrated in this figure, ${Q_h}$ is entirely independent of $\theta $.
This behavior is quite distinct from the strong $\theta $-dependence indicated by ${Q_\rho }$, implying that ${Q_h}$ is unaffected by band occupancy.
More quantitatively, the numerical results confirm a linear relation ${Q_h} = - \delta $.
Thus, the ratio ${Q_h}/\delta = -1$ is quantized independently of $\delta $ as well as $\theta $.
As shown in Figs. \ref{fig:3} (c) and (d), ${J_h}(t)$ and ${q_h}(t)$ weakly depend on $\theta $, whereas the cyclic change of ${q_h}(t)$ by the amount of ${Q_h} = - \delta $ is retained irrespective of $\theta $.

\begin{figure}
\centering
\includegraphics[width=\linewidth]{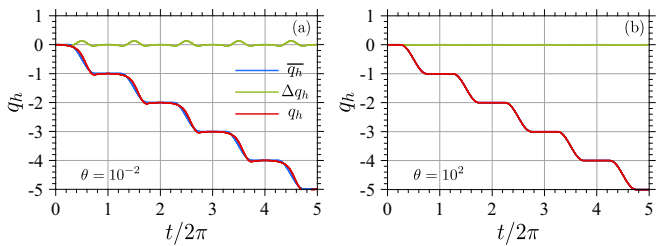}
\caption{(Color online) Time-integrated energy current ${q_h}(t)$ and its components ${{\bar q}_h}(t)$ and $\Delta {q_h}(t)$ calculated for (a) $\theta = {10^{-2}}$ and (b) $\theta = {10^2}$ in five pumping cycles. In these calculations, we assumed $\delta = 1$.}
\label{fig:4}
\end{figure}

To quantitatively understand these observations, it is worth considering the symmetry relations for the two components of energy current defined by Eqs. (\ref{eq:26}) and (\ref{eq:27}).
In the assumed pumping protocol, ${\bar J_h}(t) = {\bar J_h}( - t)$ is $t$-even, while $\Delta {J_h}(t) = - \Delta {J_h}( - t)$ is $t$-odd.
The latter automatically leads to $\Delta {Q_h} = 0$.
The vanishing $\Delta {Q_h}$ is confirmed in Fig. \ref{fig:4} where ${{\bar q}_h}(t)$ and $\Delta {q_h}(t)$ are shown separately.
Accordingly, one easily derives
\begin{equation}
\label{eq:30}
{Q_h} = {\bar Q_h} = - \delta ,
\end{equation}
from Eq. (\ref{eq:28}), since $\nu = 1$ for $\tfrac{\pi }{2} < t < \tfrac{{3\pi }}{2}$ and $\nu = 0$ otherwise.
Thus, energy pumping and its quantization are explained in terms of the global Berry phase for all bands and the topological phase transitions during the pumping cycle.

Recall that Eq. (\ref{eq:30}) assumes the symmetry $\Delta {J_h}(t) = - \Delta {J_h}( - t)$.
Thus, strictly speaking, how ${Q_h}$ varies with temperature depends on the protocol under consideration.
However, it should be emphasized that the high-temperature formula, Eq. (\ref{eq:29}), is valid for an arbitrary protocol.

\begin{figure}
\centering
\includegraphics[width=\linewidth]{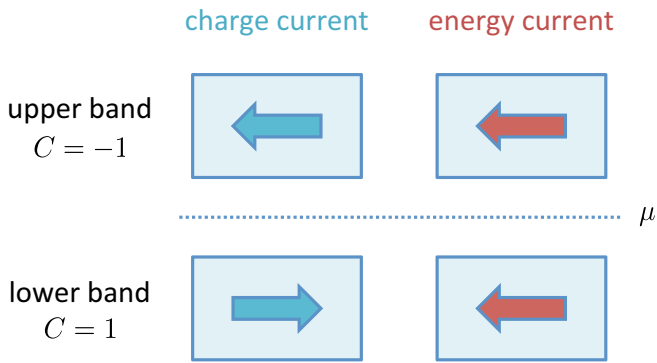}
\caption{(Color online) Sketch of charge and energy currents flowing in lower and upper bands during a nontrivial cycle.}
\label{fig:5}
\end{figure}

To intuitively derive a physical implication for energy pumping, it may be worth considering the time evolution of Wannier states during the pumping cycle \cite{ref:20,ref:21}.
In the nontrivial cycle, the center positions of Wannier states of lower and upper bands move to the neighboring cells at opposite sides, reflecting the opposite Chern numbers $ \pm 1$ of these bands.
The resulting charge and energy currents in the two bands are explained in Fig. \ref{fig:5}.
As seen from the figure, the net charge current is produced when the two bands are unequally occupied.
The counter translations of Wannier centers also bring about energy current.
Since single-particle energies have opposite signs in the two bands, energy is transported via these bands in the same direction.
This implies that, unlike charge pumping, unequal band occupancy is not required for energy pumping, putting a qualitative interpretation on ${Q_h}$ and its high-temperature behavior.

Thus far, we have assumed the half-filled system ($\mu = 0$).
Finally, we briefly summarize the $\mu$-dependence.
For $\mu = \infty$, ${f_ + } = {f_ - } = 1$ at zero and finite temperatures.
In this case, we find that ${Q_\rho } = 0$ and ${Q_h} = - 2\delta$ for the assumed protocol.
The former is obvious from ${\sum _n}{C_n} = 0$.
The latter is twice as much as ${Q_h}$ at half-filling, and is reasonably interpreted as cooperative contributions from two filled bands.
Needless to say, ${Q_\rho } = {Q_h} = 0$ for $\mu  =  - \infty $.

\section{Summary}
\label{sec:6}

We have studied energy polarization and energy pumping in the fermionic RM model at half filling.
For this model, the summation of Berry phases over all bands defines the global Berry phase and the associated winding number.
This is a $\mathbb{Z}$ invariant that identifies topological phases in generic 1D systems even in the absence of symmetry.
We analyze charge and energy polarizations using the second-quantization formalism and connect them to the relevant cell-to-cell currents in terms of current continuity equations.
The expectation values of polarizations at finite temperatures lead to generic pumping formulas that describe instantaneous currents flowing in an arbitrary adiabatic cycle.
Following these formulations, we compare charge pumping and energy pumping in a particular cycle through numerical calculations.
The pumped charge ${Q_\rho }$ per cycle is quantized at unity in the low-temperature regime, whereas it vanishes in the high-temperature limit.
Conversely, the pumped energy ${Q_h}$ is entirely temperature-independent and follows a relation ${Q_h}/\delta = - 1$ irrespective of the pumping amplitude $\delta $.
An explicit explanation is given for the energy pumping in terms of the global Berry phase for all bands and the related topological phase transitions occurring in the pumping sequence.

Energy and heat currents are equivalent at half-filling.
Hence, it is expected that a controlled cyclic modulation of parameters in a finite segment of the system produces a temperature difference between two outer portions.
This is a topological Peltier-like effect, which may be experimentally tested on optical lattices.

In the present study, we do not explicitly consider dissipation and relaxation processes, which tend to force the system into an instantaneous equilibrium state.
They could be caused by a heat bath in contact with the system.
Thus, incorporating the system-bath coupling into a concrete pumping model may be an interesting topic of future study.

\bibliography{ref}
 
\end{document}